\documentstyle[12pt,aaspp]{article}
\singlespace
\tighten
\onecolumn
\received{}
\accepted{}
\begin{document}

\def\be{\begin{equation}}
\def\ee{\end{equation}}
\def \and{\& }
\def \etal{{\it et al. }}
\def \DC{dark cluster}
\def \solarmass{M_\odot}
\def \LO85{Lacey \and Ostriker 1985}
\def \CL87{Carr \and Lacey 1987}
\def \rDC{r_{C}}
\def \MDC{M_{C}}
\def \tauDC{\tau_{C}}
\def \Vc{V_{\infty}}
\def \fC{f}
\def \rE{r_{E}}
\def \dLMC{d_{LMC}}
\def \pc{\rm pc}
\def \degsq{{\rm deg}^2}
\def \Paczy{Paczy{\'n}ski }
\def \paczy{Paczy{\'n}ski }

\title{Gravitational Microlensing By Dark Clusters In the\\
  Galactic Halo}

\vspace{0.7in}
\author{Eyal Maoz}
\vspace{0.7in}

\affil{Harvard-Smithsonian Center for Astrophysics, \\
MS 51, 60 Garden Street, Cambridge, MA 02138}

\vspace{0.3in}
\centerline{E-mail: maoz@cfa.harvard.edu}
\newpage
\begin{abstract}
The dark matter in Galactic halos, or some fraction of it, may be in the form
of dark clusters which consist of small mass objects.
Carr \and Lacey (1987) have derived the permissible properties of such
systems, and proposed the existence of dark clusters with
mass of order $\sim\!10^6\solarmass$ to explain some of the observed dynamical
properties of the stellar disk of the Galaxy.
A population of bound systems with mass of $\sim\!10^5\hbox{-}10^6\solarmass$
is also an attractive possibility since it is close to the baryon Jeans mass at
recombination, which may be the preferred
mass scale for the first bound objects to form in the universe.
At the present, the existence of dark clusters which consist
of brown dwarfs, Jupiters, or black hole remnants of an early
generation of stars, is not indicated, nor
can be excluded on observational grounds.

We describe how dark clusters can be discovered in a
sample of gravitational microlensing events in LMC stars.
Alternatively, it could provide
strict bounds on the fraction of halo mass which resides in such systems.
If MACHOs are clustered, the implied degeneracy in their spatial and velocity
distributions would result in a strong
autocorrelation in the sky position of microlensing
events on an angular scale $\lesssim\!20$
arcsec, along with a correlation in the event duration.

We argue that a small number of events could be enough to indicate
the existence of clusters, and demonstrate that a sample of $\simeq\!10$
events would be sufficient to reject the proposal of Carr
\and Lacey (1987) at the $95\%$ confidence level. If the mass of the
hypothesized clusters is much lower than $10^6\solarmass$, or the fraction
of the dark matter which resides in clusters is small, then a much
larger sample of events may be required.
\end{abstract}

\newpage

\section{INTRODUCTION}
It has been proposed that the
dark matter in galactic halos may consist of objects with mass of order
$\sim \! 10^6 \solarmass$, either in the form of
supermassive black holes
(Lacy \and Ostriker 1985; Ipser \and Semenzato 1985, but see Moore 1993), or
as dark stellar dynamical clusters of small mass objects (\CL87).
The main attraction in a population of such massive objects
is that it could naturally explain some of the observed
dynamical properties of the disk component of our Galaxy (\LO85).
This includes the observed
amount of heating of the stellar disk and its dependence on time; the
shape of the stellar velocity ellipsoid, and the
observed tail of high velocity stars in the solar neighborhood.
As pointed out by Carr \and Lacey (1987), the hypothesized clusters have some
advantages over the black holes scenario. They will not
radiate with a considerable luminosity due to gas
accretion when passing through
the Galactic disk (Carr 1979; \LO85), and will avoid a too rapid orbital
decay and accumulation in the Galactic nucleus (cluster mass loss due to
tidal disruption at small Galactocentric distances will reduce the
dynamical friction).

Assuming that most of the halo mass resides in
clusters, Carr and Lacey (1987) derived the
permissible cluster properties which can
lead to a dynamically consistent picture.
They found that the typical cluster radius is
constrained to be around $\approx \! 1\pc$ from
collisional and tidal cluster disruption considerations, and that the clusters
would have to be composed of objects with typical
mass smaller than $10\solarmass$ in order
to avoid evaporating on a timescale shorter than the age of the Galaxy.
Such clusters are unlikely to be predominantly composed of neutron stars
or white dwarfs (Boughn, Saulson, \and Seldner 1981; Gilmore \and
Hewett 1983, but see also Ryu, Olive, \and Silk 1990; Eichler \and Silk 1992),
but they may consist of Jupiters, brown dwarfs, or small black hole
remnants from an early generation of stars.

One must admit that, although shown to be a viable possibility, the idea that
the entire galactic dark matter distribution resides in dense
compact clusters may not be very appealing.
However, there is absolutely no reason to exclude the possibility that
some {\it fraction} of the halo dark mass
is in the form of $10^5\hbox{-}10^6\solarmass$ clusters.
This is not an outrageous idea since we already know
that (luminous) matter is clustered on
various scales. Also, this mass range is close to
the baryon Jeans mass at recombination, which may
be the preferred mass scale for the first bound objects to form in
the universe (Peebles \and Dicke 1968; Carr \and Rees 1984).

Recent detections of gravitational microlensing events
(Alcock \etal 1993; Aubourg \etal 1993; Udalski \etal 1993, 1994) indeed
suggest that at least some of the Galactic dark matter is in the form of
Massive Compact Halo objects (MACHOs) with masses
of order $\sim\!0.1 \solarmass$ (see also \Paczy 1986,1991; Griest 1991;
Griest \etal 1991).
In this Letter we draw the attention to
a simple test which can either reveal the existence of dark
clusters in the Galactic halo, or place strict bounds on the fraction of dark
matter which resides in MACHO clusters. We describe
the signature of clusters in a sample of microlensing events in the
direction of the Large Magellanic Cloud (\S{2}), and conclude in \S{3}.

\section{MICROLENSING BY CLUSTERS OF MACHOS}
Let us assume that a fraction $\fC$ of the Galactic
halo dark matter ($0\!\le\fC\!\le
\!1$) consists of clusters with mass $\MDC$ and radius $\rDC$, each composed
of objects with typical mass $m$.
Carr \and Lacey (1987) have demonstrated that the combination of
$\MDC\!\simeq\!
10^6\solarmass$, $\rDC\!\simeq\!1\pc$, $\fC\!=\!1$, and $m\!=0.1\solarmass$
satisfies the conceivable dynamical constraints (but see Moore 1993), so we
shall adopt these numbers as reference values.
First we derive a few
characteristics of the cluster population, and then discuss the
observational  implications on gravitational microlensing experiments.

The mass density profile of the Galactic halo in the range of Galactocentric
distances $R_0\!\le\! R\! \le\! R_{LMC}$ is well fit by
$\rho(R)\! =\! \Vc^2/(4\pi G R^2)$, where $\Vc$ is the asymptotic rotational
velocity of the disk, and $R_0 (R_{LMC})$ are the distances of the Sun (LMC)
to the Galaxy center, respectively.
The number density of clusters as a function of distance from the Sun, $r$,
in the LMC direction is
\be n(r) = {\fC \, \Vc^2 \over 4\pi G \MDC \, (
    R_0^2 + r^2 - 2R_0 r \cos\theta_{LMC} ) }  \:\:\:\: , \ee
where $\theta_{LMC}$ is the angle between the LMC and the Galactic center.
Denoting the total angular area of the LMC which is monitored for
microlensing events by $\Delta\Omega$ ($\simeq\!2\, \degsq$), the
number of foreground clusters with lines of sight which intersect that area is
\be  N \simeq \Delta\Omega \! \int_0^{\dLMC} {\!\! n(r) \, r^2 dr} \, =
    \: 22 \, \fC \left({\MDC \over 10^6 \solarmass}\right)^{-1} \!\!
          \left({\Delta\Omega \over 2 \,\degsq}\right)  \:\:\: , \ee
where we  have substituted $R_0 \!=\! 8.5$ kpc, $\Vc \!=\! 220 km s^{-1}$,
$\dLMC\!=\!55$kpc, and $\theta_{LMC}=82\deg$.
The fraction of the LMC area which is behind dark clusters is
\be   P_C \simeq\:   \Delta\Omega \int_0^{\dLMC} {\! {\pi \rDC^2
          \over  \Delta\Omega\, r^2}\: n(r)\,
               r^2 dr} \, \simeq \: 5 \times 10^{-4} \,\fC
           \left({\MDC \over 10^6 \solarmass}\right)^{-1} \!
          \left({\rDC \over 1\pc }\right)^2  \:\:\:\; .            \ee

The optical depth for gravitational microlensing,
averaged over all LMC lines of sight,
is $\approx\!5\!\times\! 10^{-7}$ (\Paczy 1986), regardless of possible
MACHO clustering.
But, the probability for microlensing of an
LMC star which is observed through a dark cluster is
\be \tauDC \approx \,  {\MDC \,\rE^2 \over m\, \rDC^2}
       \:\: + \: 5\!\times\! 10^{-7} \! \left({1 - \fC}\right) \:
   \simeq \: 1.5\times 10^{-3} \left({\MDC\over 10^6 \solarmass}\right)
   \left({\rDC\over 1\pc}\right)^{-2} \ee
where the r.h.s of equation (5) is an evaluation for a cluster at a distance
of $r\!=\!10$kpc, and the Einstein Radius, $\rE$, is defined by
\be \rE^2 = {4 G m\,  r(\dLMC -r) \over c^2 \, d_{LMC} } \:\: .   \ee

If a substantial fraction of the MACHO population resides in clusters
then the angular distribution of microlensing events in LMC stars should be
strongly correlated on very small angular scales.
Equation (3) shows that $P_C\!\ll\!1$ for a considerable
range of parameters, which
means that only a small fraction of the LMC area is observed through
clusters. Therefore, in the limit $\fC\!\rightarrow\! 1$ the
events will appear to be concentrated in $\sim\!N$ isolated small regions
(hereafter, ``spots''), where $N$ is given by equation (2). The typical
angular size of each spot would be
\be \theta_C \approx {\rDC \over 10 {\rm kpc}} =\,
            20\,arcsec\, \left({\rDC \over 1\pc}\right) \:\:\: .  \ee
The number of spots would be effectively smaller then estimated
by Eq. (2), and the reason for that is the following:
the optical depth for microlensing is proportional
to $\rE^2$, so
the probability for lensing through a cluster at a distance $r$ scales
as $r(\dLMC - r)$. This means that most events are likely to be
produced by those clusters which are located
within a (broad) range of distances
around $\simeq\! 10$kpc where $n(r)r(\dLMC-r)$ has a maximum.
Thus,
the effective number of spots is smaller than the number of clusters, $N$.

Let us estimate the number of events which are required in order to confirm or
rule out the population of $10^6\solarmass$ dark clusters ($\fC\!=\!1$) which
has been proposed by Carr and Lacey (1987).
Assuming that the monitored LMC stars are observed through a population of
$N$ clusters, the
probability that no repetition from the same cluster will occur in a sample
of $k$ events (the ``Birthday Problem'') is given by
\be q_k \simeq {N! \over (N-k)! \,  N^k} \, \simeq  \, e^{-k}
      \left({N\over N-k}\right)^{N-k+0.5} \:\:\: , \ee
where all clusters are assumed to have identical optical depth, and
the rightmost approximation is due to Stirling's formula.
Substituting $N\!=\!22$ as implied from Eq. (2) and using Eq. (7), we find
that if no pair of events
is observed at a separation $\lesssim\!\theta_C$, then a sample of $10$
events is sufficient for rejecting the proposal of
Carr and Lacey (1987) at the $\simeq\!90\%$ confidence level.
Using Monte Carlo simulations which take into account
the varying optical depth of clusters with distance, we found that the
absence of a very close pair in a sample of $10$ events enables rejecting
the above hypothesis at the
$\simeq\!95\%$ confidence level
(the higher confidence level reflects the fact that
the effective number of spots is smaller than $N$, as discussed earlier).
Introducing Poisson noise in $N$ into the simulations resulted in a negligible
change in the number of required events (the contribution of realizations
with $N\!\gtrsim\!22$ compensates almost entirely that of ones with
$N\!\lesssim\!22$).

At the present,
the closest pair among the 3 events observed towards the LMC
(Alcock \etal 1993;
Aubourg \etal 1993) is at a separation of $2.38\deg$ (this does not conflict
with the fact that the total monitored LMC area in the MACHO experiment is
$2\,\degsq$ since it is composed of four different fields).
But, if a pair with angular separation
$\lesssim\!\theta_C$ will be found in a sample of a few
events, and the probability that this occurs by chance when there is no
clustering is very small,
then it may strongly suggest that dark clusters do exist.

In order to rule out a population of lower mass clusters (still with
$\fC\!=\!1$) the number of observed events should be of order
$\approx\! 10 (\MDC/10^6\solarmass)^{-1/2}$. This results from the
scaling $k\!\propto\! N^{1/2}$ for a fixed value of $q_k$, normalized
to $10$ events for clusters with mass of $10^6\solarmass$. One should bear
in mind that a low value of $\MDC$ implies a higher number of clusters, so
a pair of events at angular separation $\lesssim\!\theta_c$ could appear by
chance, in which case the above test would
not be adequate.
When a large sample of events is established, the only
reliable way to evaluate the statistical significance for ruling out various
portions of the ($\fC,\MDC,\rDC$) parameter
space would be using Monte Carlo simulations. These
could also take the various observational biases,
as well as the distribution of monitored LMC stars, into account.

A necessary condition for observing repeated events from the same cluster is
that the average number of {\it monitored\/} LMC stars
behind a cluster be more than one.
Roughly $1.8$ million LMC stars are currently being monitored in the MACHO
project for
brightness variations within a total area of $\simeq\!2\,\degsq$
(K. Griest, private communication), so
their average number behind each cluster is
\be \sim\, 1.8\!\times\!10^6 \,\left({\pi \theta_C^2 \over\Delta\Omega}\right)
 \simeq \: 10^2  \,\left({\rDC\over 1\pc}\right)^2            \:\:\: , \ee
and the condition is satisfied.

An additional and independent test for the existence of dark clusters is
provided by the distribution of event durations.
The time scale for the intensity variation of a microlensed star depends
on the lens's mass and distance, and on the relative tangential velocity
between the star and the lens. Since each of these
three quantities may vary at least by an order of magnitude for Galactic
halo objects,
the distribution of event durations is expected to be quite broad.
However, if events are produced by
MACHOs within the same cluster, then the corresponding lenses
are practically at identical distances, and have roughly the same velocity
vector up to a correction of order $\sigma_C/\Vc$, where $\sigma_C$
is the velocity dispersion of MACHOs within a cluster. In such case,
the dispersion in the distribution of relative tangential
velocities between a lens and a source will be reduced by a factor of
$\left[{(\sigma_C^2+\sigma_{LMC}^2)/(\Vc^2+\sigma_{LMC}^2)}\right]^{1/2}
\!< 1$, where $\sigma_{LMC}$ is the velocity dispersion of the LMC stars.
Therefore, the dispersion in event duration within each spot
will be mainly due to
the dispersion in the lens' masses, and events which are correlated in
sky position should also be correlated in duration. When a sufficiently
large sample of events is established,  the dispersion
in event duration within each ``spot'' should be significantly lower than
the dispersion in the entire sample.

Two important characteristics of a microlensing event are that it
may occur
(practically) only once for a given star, and that the light curve
be symmetric around the maximum of brightness and have a characteristic shape
(\Paczy 1991).
An exception for the later is when either the
lens or the source is a binary system or
has a planet (Mao \and \Paczy 1991; Gould \and Loeb 1992).
If the hypothesized clusters had been dense enough, the projected
separation between MACHOs could have been comparable or smaller than the
Einstein radius. In such case the light curve would not be symmetric, and
stars would be observed to undergo repeated events. Equation (4)
shows that this is unlikely to happen since the mean optical depth for
microlensing through a cluster ($\sim\!10^{-3}$) is considerably smaller
then unity for a wide range of cluster properties (this is independent of
the number of MACHOs per cluster). Nevertheless, assuming a typical MACHO mass
of order $\sim\!0.1\solarmass$, and that the central projected density of a
cluster may be an order of magnitude larger than the mean one, we may expect
a few particular stars which happen to be aligned with a cluster center
to exhibit repeated microlensing events (and more complex light curves) on a
timescale of $\sim\!10^2$ months.


In the future, if the existence of dark clusters is established, then
a large sample of microlensing events could
provide invaluable information on the projected
structure of the clusters, their tangential velocities, and
the mass distribution of MACHOs.

\section{DISCUSSION}
We described an {\it observational\/} test which can provide either
an indication or constraints on the existence of dark clusters in the
Galactic halo, using a sample of gravitational microlensing events in the
direction of the LMC.
We have shown that valuable information could be
extracted already from a sample of a few events.
For example, roughly 10 events in the LMC direction could be sufficient to
rule out the population of $\simeq\!10^6\solarmass$ clusters which has been
proposed by Carr and Lacey (1987). On the other hand, if a comparable number
of events exhibit clustering in sky position on scales $\lesssim\!20$ arcsec,
as well as the expected correlation in the event duration (\S{2}),
it would strongly suggest that clusters do exist.

Gould (1993) has shown that if the LMC acquires a halo of MACHOs, it would
produce a variation of the optical depth to lensing as a function of
sky position in the LMC
by as much as $\sim\!20\%$. This will not affect the
correlation of events on very small angular separations, but it should
be taken into account in the future when a large sample of events is
established and confronted with Monte Carlo Simulations.

The test for MACHO clustering could in principle be applied also to
microlensing of stars in the Galactic bulge where four events have already
been observed (Udalski \etal 1993, 1994).
However, although interesting by itself,
it would not provide much valuable information on dark clusters
since those are likely to be disrupted at small Galactocentric distances
(i.e., $\lesssim\!R_{0}/2$, where the optical depth for lensing in that
direction is the largest). On the other hand, if clustering is observed in
that direction it may be explained, for example, by a small scale
clustering of protostars in the Galactic disk.

We assumed that the observed microlensing events are produced by MACHOs
in the Galactic halo. However, it is still possible that the
observed events are due to MACHOs
which are distributed in a thick disk rather than in the halo, in which case
the test described in this Letter is inadequate.
A larger sample of events will enable to discriminate among these
two possibilities (Gould, Miralda-Escude, \and Bahcall 1994; Gould 1993).

I wish to thank  Rosanne Di Stefano, Chris Kochanek, Shude Mao,
and Martin Rees for discussions and comments.
This work was supported by the U.S. National Science Foundation, grant
PHY-91-06678.

\end{document}